\documentstyle[prl,aps,multicol,graphicx,amssymb,latexsym]{revtex}

\newcommand{\be}{\begin{equation}} \newcommand{\ee}{\end{equation}}

\newcommand{\ba}{\begin{eqnarray}}
\newcommand{\ea}{\end{eqnarray}}

\begin{document}

\title{World-Line Path Integral Study of \\Supersymmetry Breaking in the Wess-Zumino Model}

\author{M. Beccaria${}^{a,b}$ and C. Rampino${}^a$}

\address{${}^a$ Dipartimento di Fisica dell'Universit\`a di Lecce,
I-73100, Italy,\\${}^b$Istituto Nazionale di Fisica Nucleare - INFN,
Sezione di Lecce }

\maketitle

\begin{abstract}
We study supersymmetry breaking in the lattice N=1 Wess-Zumino model 
by the world-line path integral algorithm. 
The ground state energy and supersymmetric Ward identities are exploited 
to support the expected symmetry breaking in finite volume.
Non-Gaussian fluctuations of the topological charge are discussed and related to the
infinite volume transition.
\end{abstract}


\begin{multicols}{2}
\narrowtext

An important issue in the study of lattice supersymmetry is the analysis of the phase transition
associated to spontaneous supersymmetry breaking~\cite{SUSYLattice}. 
The problem is particularly difficult
in 1+1 dimensions where 
non perturbative effects can be dominant~\cite{Grisaru}.
In this Report, we address the determination of the phase diagram in the N=1 Wess-Zumino (${\rm WZ}_{1}$) model.
The transition must be analyzed by accurate sampling of the ground state and, 
to this purpose, there exist well established specific
numerical techniques, {\it e.g.}
Green Function Monte Carlo (GFMC)~\cite{GFMC}.
In a recent series of papers~\cite{WZus}, the model ${\rm WZ}_1$  has been studied by 
GFMC with adaptive trial wave function optimization.
The main source of systematic error turns out to be the finite number $K$ of quantum states that
represent stochastically the ground state. The $K\to\infty$ extrapolation is delicate 
because the analytical form of the asymptotic corrections is not known. 
For these reasons, it is worth exploring alternative methods like 
the so-called world-line path integral (WLPI)~\cite{WLPI},
more common in the study of high energy lattice models. 
In the framework of WLPI quantum expectation values are
evaluated at finite temperature $T=1/\beta$ and the ground state projection 
($\beta\to\infty$) is numerically estimated.
In one spatial dimension, fermion configurations are sampled by  
local moves of the  fermion lattice world lines. Previous WLPI studies of 
${\rm WZ}_1$ considered small lattices at moderately low temperatures~\cite{RanftSchiller}.
A signal for the transition was given in terms of a small ground state energy compatible with zero 
within the achieved precision. This interpretation is problematic since analytical 
arguments~\cite{Witten} as well as GFMC data~\cite{WZus} support SUSY breaking in finite 
volume for the considered model.
A first aim of this Report is that of clarifying the results obtained with the WLPI
method. We extend the analysis by including SUSY Ward identities 
and discussing  the removal of systematic errors on larger lattices and lower temperatures.
The topological charge is also measured and  non-Gaussian fluctuations
are found. Recently, the fluctuation spectra  of global observables has been proposed
as a monitor for criticality even relatively far away the transition~\cite{Universal}.
Indeed, we shall motivate the conjecture that the topological charge plays this role
in the ${\rm WZ}_1$ model.

The $(1,1)$ SUSY algebra in two dimensions has two fermionic generators $Q_{1,2}$
and three bosonic ones: the components of the two-momentum $(P^0, P^1)$
and a central charge $Z$. The non trivial algebra is 
\be
\{Q_a, Q_b\} = 2(P^0\ {\bf 1} + P^1\sigma^1 + Z\sigma^3)_{ab},
\ee
where $\sigma^i$ are Pauli matrices.
The ${\rm WZ}_1$ model realizes the above algebra on a real chiral multiplet with a real scalar component 
$\varphi$ and a Majorana fermion with components $\psi_{1,2}$. The supercharges are
\be
Q_{1,2} = \int dx 
\left[p\ \psi_{1,2}
-\left(\frac{\partial\varphi}{\partial x}\pm V(\varphi)\right)\psi_{2,1}\right] ,
\ee
where $p(x)$ is the momentum operator conjugate to $\varphi(x)$
and $V(\varphi)$ is an arbitrary function called prepotential in the following.
The central charge is a topological quantum number~\cite{WittenOlive} related to the 
existence of soliton sectors.
On the lattice there are no continuous translation and we preserve a 
SUSY subalgebra~\cite{Elitzur}.
We choose one of the supercharges, say $Q_1$, build a discretized version $Q$ and 
define the lattice Hamiltonian to be 
$H = Q^2$.
By Monte Carlo, we look for the eigenspace of $Q^2$ with lowest eigenvalue. The simulation 
explores all the topological sectors: the $Q$-invariant states with $Z=0$ are 
expected to be lattice versions of fully supersymmetric ground states. In principle, 
there are also $Q$-invariant states with $Z\neq 0$, {\it i.e.} BPS states~\cite{WittenOlive}. 
The transition between an empty
$Q^2=0$ eigenspace and a non trivial one is a weak (actually halved) definition of SUSY breaking 
that we shall adopt in this Report~\cite{WZus}.

\begin{figure}[htb]
\centerline{\includegraphics*[width=8cm,angle=0]{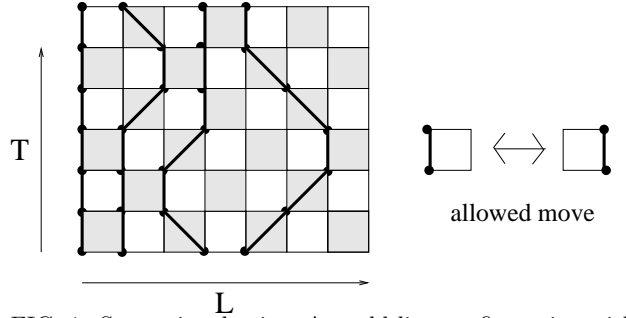}}
\caption{Space time lattice. A world-line configuration with 4 fermions is shown.}
\label{fig:1}
\end{figure}

The explicit lattice model is built by considering a spatial
lattice with $L$ sites and open boundary conditions. On each site we assign
a real scalar field $\varphi_n$ with its conjugate momentum $p_n$ such 
that $[p_n, \varphi_m] = -i\delta_{n,m}$.  The associated fermion is a Majorana fermion
$\psi_{a, n}$ with $a=1, 2$ and $\{\psi_{a, n}, \psi_{b, m}\} =
\delta_{a,b}\delta_{n,m}$ , $\psi_{a,n}^\dagger = \psi_{a,n}$. The
discretized fermionic charge and lattice (central) topological charge are
\ba
Q &=& \sum_{n=1}^L\left [
p_n\psi_{1,n}-\left(\frac{\varphi_{n+1}-\varphi_{n-1}}{2}
+V(\varphi_n)\right)\psi_{2,n}\right] , 
\ea
\be
Z = \sum_{n=1}^L\left[ \frac{\varphi_{n+1}-\varphi_{n-1}}{2}\ V(\varphi_n) \right] .
\ee
For a smooth configuration the continuum limit of $Z$ is a surface term.
Following \cite{RanftSchiller} we replace the two Majorana fermion operators
with a single Dirac operator $c$ satisfying canonical
anticommutation rules, {\it i.e.},
$\{c_n, c_m\} = 0$, $\{c^\dagger_n, c^\dagger_m\} = 0$, $\{c_n,c_m^\dagger\} = \delta_{n,m}$:
\be
\psi_{\{1,2\},n} = \frac{(-1)^n \mp i}{2i^n}(c_n^\dagger\pm ic_n) .
\ee
The Hamiltonian is $H \equiv Q^2 = H_B+H_F$ with 
\ba
H_B &=& \sum_{n=1}^L\left\{ \frac 1 2\ p_n^2 + \frac 1
2\left(\frac{\varphi_{n+1}-\varphi_{n-1}}{2} + V(\varphi_n)\right)^2
\right\} , \\
H_F &=& \sum_{n=1}^L\left\{-\frac 1 2\ (c^\dagger_n c_{n+1} + c^\dagger_{n+1}c_n) + \right. \\
&& \qquad\qquad\qquad \left . +(-1)^n V'(\varphi_n) \left(c^\dagger_nc_n-\frac 1 2 \right)
\right\} . \nonumber
\ea
$H$ conserves the total fermion number
\be
N_f = \sum_{n=1}^L N_n,\qquad N_n = c^\dagger_nc_n ,
\ee
and can be examined separately in each sector with fixed $N_f$.
The simplest observable that we consider is the lattice ground state energy 
$E_0 = \min\mbox{spec} (Q^2) \ge 0$.
Additional quantities are related to global supersymmetric Ward identities as follows.
If the vacuum $|\Omega\rangle$ obeys $Q|\Omega\rangle=0$, for any operator $A$ we have 
$\langle \Omega |\ \{Q, A\}\ | \Omega\rangle = 0$.
In particular, taking $A = \sum_{n=1}^L F(\varphi_n)\ \psi_{2, n}$, we obtain
\ba
W = \langle 0 | \sum_{n=1}^L\left\{ \right. && 
\left .F(\varphi_n)\left[\frac {\varphi_{n+1}-\varphi_{n-1}}{2} + V(\varphi_n)
\right] + \right .\\ 
&&\left .+ F'(\varphi_n) (-1)^n (c_n^\dagger c_n -1/2)\right\}
| 0\rangle = 0 . \nonumber
\ea
A set of  independent Ward Identities is obtained by considering 
$F(\varphi) = \varphi^k$. We shall call $W_k$ the corresponding quantities.
In summary, $Q|\Omega\rangle=0\leftrightarrow E_0=0$ and also  $Q|\Omega\rangle=0\rightarrow W_k=0$.
The world-line path integral computes the expectation value of an operator $A$ 
at inverse temperature $\beta$ as
$\mbox{Tr}\ (A\ e^{-\beta H})/\mbox{Tr}\ e^{-\beta H}$.
As usual the trace is computed by Trotter splitting of $e^{-\beta H}$ with time spacing 
$\varepsilon = \beta/T$
\be
\label{Trotter}
e^{-\beta H} = (e^{-\varepsilon H_B}e^{-\varepsilon H_F^e}e^{-\varepsilon\ H_F^o})^T + 
{\cal O}(\varepsilon^2)
\ee
where $H_F^{e(o)} = \sum_{n\ \mbox{even} (\mbox{odd})} H_{F,n}$,
\ba
H_{F,n} &=& -\frac 1 2 (c^\dagger_n c_{n+1} + {\rm h.c.}) + 
\zeta_n (-1)^n V'(\varphi_n) \left(N_n-\frac 1 2 \right)  \nonumber\\
&& + \zeta_{n+1} (-1)^{n+1} V'(\varphi_{n+1}) \left(N_{n+1}-\frac 1 2 \right),
\ea
with $\zeta_n=1$ on sites $0$ and $L-1$ and $1/2$ elsewhere. The terms $H_{F,n}$ with 
even $n$ commute among themselves as well as the terms with odd $n$. 
Complete sets of intermediate states are inserted between the various factors in Eq.~(\ref{Trotter}).
The space-time lattice is shown in Fig.~(\ref{fig:1}). The weight of a fermion configuration is computed by 
multiplying the weights associated to the shaded squared plaquettes. 
The finite temperature expectation values of the energy, $Z$ and the 
Ward identities can be readily measured. We choose to 
evaluate the kinetic bosonic energy by the Virial theorem.
A full sweep is a heat bath on the unshaded plaquettes with the elementary moves shown 
in Fig.~(\ref{fig:1}) followed by a multi-hit Metropolis update on the field $\varphi$ configuration.
A detailed discussion of such algorithms can be found, {\it e.g.}, in~\cite{WLPI}.

\begin{figure}[htb]
\centerline{\includegraphics*[width=8cm]{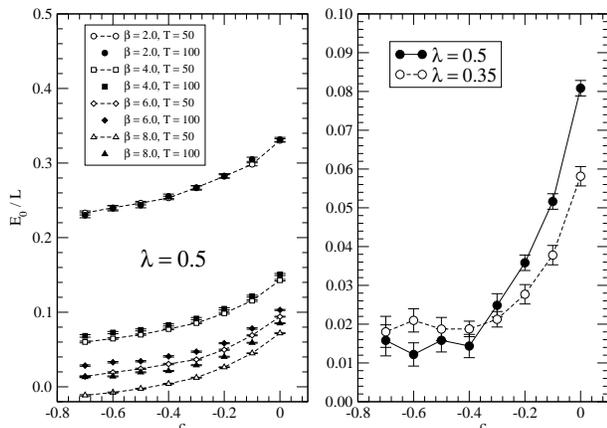}}
\caption{Ground state energy per site $E_0/L$. The left graph shows the intermediate results at variable
$\varepsilon$ and $\beta$. The right graph shows the extrapolation to $\epsilon\to 0$ followed by 
$\beta\to\infty$.}
\label{fig:2}
\end{figure}

For the numerical analysis, we focus on the specific
prepotential $V(\varphi) = \lambda \varphi^2 + c$
that despite its simplicity already shows all the interesting features of the symmetry breaking. 
At the classical
level, SUSY is broken for $c>0$. For negative $c$, the zeroes of $V$ are associated to 
classical SUSY vacua and their fate after radiative corrections must  be investigated.
For this particular $V$, the analysis of the quantum model in the continuum can be pursued in full details 
and leads to the following conclusions~\cite{Witten}.
In finite volume, tunneling effects  lift the ground state energy to 
a positive value for all (positive and negative) $c$. In infinite volume and sufficiently large $c<0$, 
there can be inequivalent vacua with $\langle \varphi\rangle\neq 0$. 
The fermion acquires a mass and the Goldstino required for spontaneous breaking is absent.
The phase diagram of the lattice model is studied in the plane $(\lambda, c)$ of adimensional couplings.
The continuum limit is obtained when $\lambda\to 0$. The Renormalization Group
trajectories with constant Physics are:
\be
\label{RG}
c -\frac{1}{2\pi}\lambda\log\lambda = \mbox{constant} .
\ee
In the scaling region, SUSY is expected to be restored below a certain critical $c$. If the
physical volume is not large enough, the transition is smoothened and a residual
positive energy remains (even for quite negative $c$) as a consequence of the above mentioned SUSY breaking
in finite volume.
The data analysis must be done  after the removal of the systematic errors
associated to $\varepsilon$ and $\beta$. Expectation values are obtained
only after the combined limit $\varepsilon\to 0$ followed by $\beta\to\infty$.
This extrapolation makes the determination of the transition a rather expensive problem. 
In the first part of our analysis, we consider the $\varepsilon$ and $\beta$ dependence of the various 
observables. To this aim, we perform explicit simulations 
with $L=22$ and the two values  $\lambda=0.35, 0.5$ that are reasonably in the scaling region~\cite{WZus}. 
We take measures at $T=50, 100$ with $2\le \beta\le 8$.
The number of full updates is typically $10^6$. We apply an high number of Metropolis hits 
($\sim 50$) in order to match the accuracy of the fermionic heat bath sampling.
We always look at the sector with $N_f=L/2$ where the ground state is expected to lie~\cite{WZus}.
In Fig.~(\ref{fig:2}) we show $E_0$ at various $\varepsilon=\beta/T$ and $\beta$; we also show the
results of a quadratic extrapolation to $\epsilon\to 0$ followed by an exponential (plus constant)
one to estimate the $\beta\to\infty$ limit.
\begin{figure}[htb]
\centerline{\includegraphics*[width=8cm]{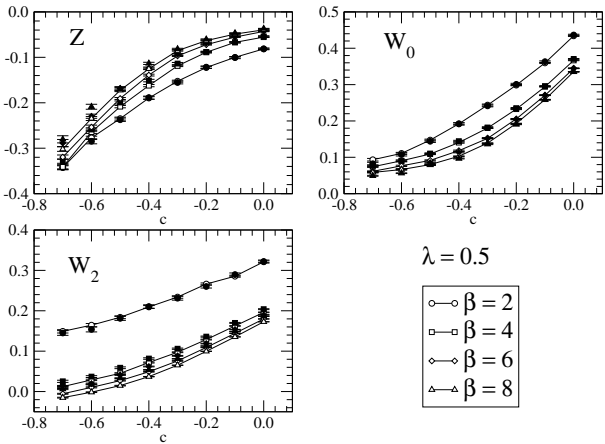}}
\caption{Intermediate results for $Z$ and $W_{0,2,4}$. Empty (full) symbols are 
obtained on the $22\times 50$ ($22\times 100$) lattice.}
\label{fig:3}
\end{figure}
The systematic error due to $\varepsilon$ increases with $\beta$ at fixed $T$ simply because
the corresponding  $\varepsilon$ is bigger. 
The limit $\varepsilon\to 0$ is absolutely necessary and, in particular, restores
positivity of $E_0$ at large $\beta$. As $c$ is decreased, the extrapolated $E_0$ reaches a 
 positive plateaux. This residual value is slightly higher at $\lambda=0.35$, due to the larger
finite volume effects. The beginning of the flat region is at $c(0.5)\simeq -0.4$ and $c(0.35)\simeq -0.3$,
in agreement with Eq.~(\ref{RG}).
In Fig.~(\ref{fig:3}) we show the intermediate data for $Z$ and  $W_{0,2}$.
It is interesting to note that $W_0$ is quite independent on $\varepsilon$ at the explored values
and is thus the most safe quantity to be analyzed. 
Fig.~(\ref{fig:4}) collects the extrapolated data.
\begin{figure}[htb]
\centerline{\includegraphics*[width=8cm]{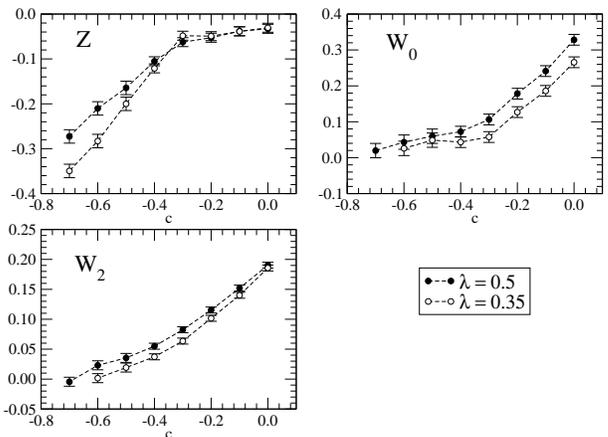}}
\caption{Extrapolated results for $Z$, $W_{0,2,4}$ obtained from the data shown in Fig.~(\ref{fig:3}).}
\label{fig:4}
\end{figure}
The central charge $Z$ is definitely negative with a sharp variation around $c\simeq -0.3$. If SUSY
is to be recovered in the $L\to\infty$ limit, we expect BPS states to be present together with eventual 
$Q=Z=0$ states. The sign of $Z$ is easily checked: in the ${\rm WZ}_1$ model we have $Q_1^2 = P^0 + Z$
and $P^0 = Q_1^2 + Q_2^2$ is a positive operator. Hence, states in the lowest $Q_1^2$ eigenspace 
prefer $Z<0$ and states with $Q_1=0$ have rigorously $Z\le 0$.
The Ward identities are all non zero although $W_0$ and $W_2$ approach smaller values
as $c$ decreases. Again $W_0$ flattens below $c\simeq -0.3$.
We conclude that SUSY is broken (at $L=22$) for all $c$ in agreement
with the GFMC analysis of~\cite{WZus} and the analytical arguments~\cite{Witten}. The SUSY 
restoration observed in~\cite{RanftSchiller} must be explained in terms of the moderate statistical errors.
Indeed, as we have shown, the absolute values of $E_0$ are quite small and arise from large cancellations
between contributions from $H_B$ and $H_F$. The search for SUSY restoration requires
a suitable finite size scaling analysis of data at rather larger $L$. The critical point is to be 
searched around $c=-0.4 (-0.3)$ for the considered $\lambda=0.5 (0.35)$. Indeed, the presented results
suggest some kind of transition there.

\begin{figure}[htb]
\centerline{\includegraphics*[width=8cm]{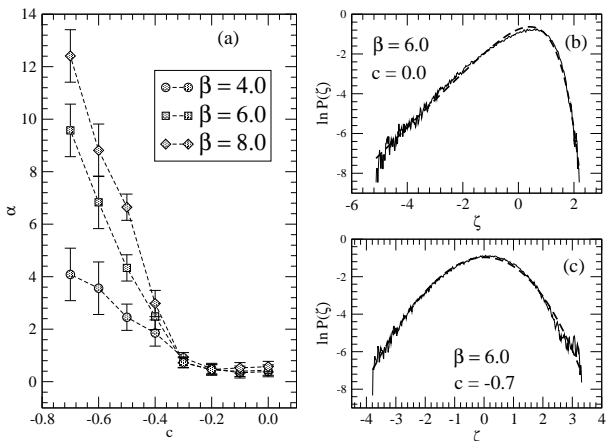}}
\caption{Analysis of $Z$ ($\lambda=0.35$). The graph (a) shows the coefficient $\alpha$ (defined in the text) as 
a function of $c$. The graphs (b,c) show two sample fluctuation spectra together with the best 
fit with Eq.~(\ref{univ}).}
\label{fig:5}
\end{figure}

In the final part of this Report we want to show how a consistent additional information  
is provided by the fluctuations of the central charge $Z$. In the continuum,
$Z$ is a global quantity depending on the boundary conditions. 
On the lattice, its geometrical meaning is not clear due to 
the quantum roughness of the configurations. 
An histogram analysis of $Z$ reveals that definite non-Gaussian fluctuations are present.
The density of the normalized variable 
$\zeta = (Z-\langle Z\rangle)/(\langle Z^2\rangle-\langle Z\rangle^2)^{1/2}$ turns out to be
consistent with the generalized Gumbel one~\cite{Gumbel}
\be
\label{univ}
P(\zeta) = {\cal N} \exp[\alpha\ (\mu\zeta+\nu-e^{\mu\zeta+\nu})],
\ee
where ${\cal N}$, $\mu$ and $\nu$ are functions of $\alpha$ fixed by normalization, zero mean and unit
variance. 
Fig.~(\ref{fig:5}) shows the best fit coefficient $\alpha(c)$ and two sample spectra.
In Fig.~(\ref{fig:5}, a)  
a change of behaviour occurs around $c^*\simeq -0.3$.
When $c\ge c^*$, $\alpha(c)$ is small and weakly dependent on both $c$ and $\beta$. Below $c^*$,
$\alpha(c)$ increases rapidly as $c$ is decreased and the slope is larger as $\beta$
increases. The values of the fitted $\mu$, $\nu$ for large $\alpha$ are such  that 
$\sqrt{\alpha}\mu\to 1$ and $\sqrt{\alpha}\nu\to 0$. For $\alpha\to\infty$, the limiting form of Eq.~(\ref{univ})
converge to a normal Gaussian. Indeed the spectrum asymmetry decreases 
as $c$ crosses $c^*$. For instance, Fig.~(\ref{fig:5}, c) is almost consistent with a simpler Gaussian fit.
Similar features are discussed in~\cite{Universal} where the particular value $\alpha=\pi/2$ is observed and 
related to the joint effect of finite-size, strong correlation and 
self-similarity. In the ${\rm WZ}_1$ model, the quantum symmetry breaking is not 
explained in terms of self-similar structures and $\alpha$ is just a  measure
of the deviation from Gaussian statistics. 
It is tempting to interpret Fig.~(\ref{fig:5}) as another signal for a
finite size critical $c^*(L)$ with 
SUSY restoration in infinite volume at $c=c^*(\infty)$.
At a qualitative level, in the symmetric phase soliton states appear with a 
definite $Z$ and the MC measures fluctuate with a normal distribution.
In the broken phase $Z$ is expected to vanish, but an inspection of the 
MC histories shows that rare configurations appear with unphysical kinks responsible for the tail of 
$P(\zeta)$ at negative $\zeta<0$.
Further analysis is needed to determine the precise functional form 
of $Z$, particularly at large $|\zeta|$, and to clarify its relation with the $L=\infty$ SUSY transition. 
However, Fig.~(\ref{fig:5},~a) demonstrates clearly that the study of $P(\zeta)$ is in principle 
rich of information.
It is a pleasure to thank M. Campostrini and A. Feo for discussions about the ${\rm WZ}_1$
model and E. Alfinito for bringing the role of 
extremal statistics to our attention.
Financial support from INFN, IS-RM42 is acknowledged.

\end{multicols}

\begin{references}

\bibitem{SUSYLattice} 
A. Feo,
{\em Supersymmetry on the lattice},
Plenary talk at 20th Int. Symposium on Lattice Field Theory (LATTICE 2002), 
Boston, Massachusetts, {\tt hep-lat}/0210015. For WZ models in 1+1 see also~\cite{RanftSchiller}
below and 
M. Beccaria, E. D'Ambrosio, G. Curci, Phys.\ Rev.\ D\ {\bf 58}, 065009 (1998);
S. Catterall, E. Gregory, Phys.\ Lett.\ B\ {\bf 487}, 349 (2000);
S. Catterall, S. Karamov, Phys.\ Rev.\ D\ {\bf 65}, 094501 (2002);
K. Fujikawa, {\em ibid.} {\bf 66}, 074510 (2002).

\bibitem{Grisaru}
L. Alvarez-Gaume, D. Z. Freedman, M. T. Grisaru,
{\em Spontaneous breakdown of supersymmetry in two-dimensions}, 
Brandeis U. preprint HUTMP 81/B111, unpublished.

\bibitem{GFMC} Recent papers exploiting GFMC methods in lattice models with no 
supersymmetry are:
M. Beccaria and A. Moro, Phys.\ Rev.\ D\ {\bf 64}, 077502 (2001);
M. Beccaria, {\em ibid.} {\bf 62}, 034510 (2000); {\bf 61}, 114503 (2000);  
Eur.\ Phys.\ J.\ C\ {\bf 13},357 (2000);
C.J. Hamer, M. Samaras, R.J. Bursill, Phys.\ Rev.\ D\ {\bf 62}, 074506 (2000);
{\bf 62}, 054511 (2000);

\bibitem{WZus}
M. Beccaria, M. Campostrini, A. Feo, 
Nucl.\ Phys.\ B\ {\bf (Proc.\ Suppl.) 106}, 944 (2002);
M. Beccaria, M. Campostrini, A. Feo, 
20th Int. Symposium on Lattice Field Theory (LATTICE 2002), 
Boston, Massachusetts, {\tt hep-lat/0209010}
to appear on Nucl.\ Phys.\ B\ {\bf  (Proc.\ Suppl.)};
M. Beccaria, M. Campostrini and A. Feo, in preparation.

\bibitem{WLPI}
The first reference to WLPI is 
J. E. Hirsch, R. L. Sugar, D. J. Scalapino, R. Blanckenbecler,
Phys.\ Rev.\ B\ {\bf 26}, 5033 (1982).


\bibitem{RanftSchiller}
J.~Ranft, A.~Schiller, Phys.\ Lett.\ B\ {\bf 138}, 166 (1984);
J.\ Phys.\ G\ {\bf 12}, 935 (1986).

\bibitem{Witten}
E.~Witten,
Nucl.\ Phys.\ B\ {\bf 202}, 253 (1982).

\bibitem{Universal}
S. T. Bramwell et al., Phys.\ Rev.\ E\ {\bf 63}, 041106 (2001);
S. T. Bramwell et al., Phys.\ Rev.\ Lett.\ {\bf 84}, 3744 (2000);
S. T. Bramwell, P. C. W. Holdsworth and J.-F. Pinton, Nature {\bf 396}, 552 (1998);

\bibitem{WittenOlive}
E. Witten and D. Olive, Phys.\ Lett.\ {\bf 78}\ B, 97 (1978).

\bibitem{Elitzur}
S.~Elitzur, E.~Rabinovici, A.~Schwimmer, 
Phys.\ Lett.\ B\ {\bf 119}, 165 (1982);
S.~Elitzur and A.~Schwimmer, Nucl.\ Phys.\ B\ {\bf 226}, 109 (1983).

\bibitem{Gumbel} E. J. Gumbel, {\em Statistics of Extremes}, Columbia
University Press, New York, (1958).
Phys. Rev. D Server 


\end{references}
\end{document}